\documentclass[12pt]{elsart}
\usepackage{graphicx}
\usepackage{color}
\usepackage{subfigure}

\usepackage{natbib}

\begin{document}
\begin{frontmatter}

\title{Observed flux density enhancement at submillimeter wavelengths 
during an X-class flare}

\author[1]{G. Cristiani\corauthref{cor}},
\ead{gcristiani@iafe.uba.ar}
\corauth[cor]{Corresponding author}
\author[2]{C.G. Gim\'enez de Castro},
\author[1]{M.L. Luoni},
\author[1]{C.H. Mandrini\thanksref{coni}},
\author[1]{M.G. Rovira\thanksref{coni}},
\author[2,3]{P. Kaufmann},
\author[4]{M. Machado}

\thanks[coni]{Member of the Carrera del Investigador Cient\'\i fico, 
        CONICET, Argentina}
\maketitle

\address[1]{Instituto de Astronom\'{\i}a y F\'{\i}sica del Espacio, CC. 67
Suc. 28, CONICET-UBA, 1428, Buenos Aires, Argentina.}
 
\address[2]{Centro de R\'adio Astronomia e Astrof\'{\i}sica
Mackenzie, Universidade Presbiteriana Mackenzie, S\~ao Paulo, Brazil.}

\address[3]{CCS, Universidade Estadual de Campinas,
Campinas, Brazil.}

\address[4]{Comisi\'on Nacional de Actividades Espaciales, Paseo Col\'on 751, 
1063, Buenos Aires, Argentina.} 

\begin{abstract}
We analyse the 30 October, 2004, X1.2/SF solar event that occurred in AR
10691 (N13 W18) at around 11:44 UT. Observations at 212 and 405 GHz of the
Solar Submillimeter Telescope (SST), with high time resolution (5 ms), show
an intense impulsive burst followed by a long-lasting thermal phase. EUV
images from the Extreme Ultraviolet Imaging Telescope (SOHO/EIT) are used to 
identify the possible emitting sources. Data from the Radio Solar Telescope 
Network (RSTN) complement our spectral observations below 15 GHz. During the 
impulsive phase the turnover frequency is above 15.4 GHz. The long-lasting 
phase is analysed in terms of thermal emission and compared with GOES 
observations. From the ratio between the two GOES soft X-ray bands, we derive 
the temperature and emission measure, which is used to estimate the free-free 
submillimeter flux density.  Good temporal agreement is found between the 
estimated and observed profiles, however the former is larger than the latter.
\end{abstract}


\end{frontmatter}

\section{Introduction}

It is widely accepted that solar flares imply the release of large amounts
of energy, which go mainly in the acceleration of particles: electrons and
ions.  The interaction of these particles with the medium, plasma and magnetic 
field, produces radiation by very different mechanisms, which is observed at 
Earth in a wide range of wavelengths. Of particular interest are electrons 
moving along the magnetic lines and producing gyrosynchrotron emission 
observed at microwaves and shorter wavelengths. Gyrosynchrotron emission of 
mildly relativistic electrons is distributed in a broad continuum at high
harmonics of the fundamental frequency $\nu_B = eB/(2\pi m_e c)$, where $e$ is 
the electron charge, $B$ the magnetic field strength, $m_e$ the electron mass 
at rest and $c$ the speed of light.  The maximum emission is centred around
$\nu\simeq\nu_B \gamma^2$, with $\gamma$ the electron Lorentz factor
\citep{Dulk:1985}. Taking $100 \le B \le 1000$ G, indicative values of~$\gamma$
 are between 2 and 5 for a 10 GHz photon. Observationally, 
\citet{Kosugietal:1988} found that the highest correlation between centimetric 
wavelengths and hard X-rays (HXR) happens between 17 GHz and 80 keV; that is, 
electrons with energies $\le 200$ keV. Therefore, radio observations at shorter
 wavelengths offer new insights to understand processes that involve high 
energy electrons during solar flares, where current HXR detector technology 
does not provide enough sensitivity and/or signal to noise ratios to observe 
weak flares. Since 1999, the only submillimeter instrument dedicated to solar 
observation is the Solar Submillimeter Telescope \citep[SST, see e.g.,
][]{Kaufmannetal:2001} operating at 212 and 405 GHz. From the theoretical 
arguments given above, $>$~4 MeV electrons should produce the synchrotron 
emission observed with SST.\\

In this work, we report the multiwavelength observations of the impulsive
and gradual phases of an X-class flare that occurred on 30 October, 2004. In 
the following section we describe the instruments and data analysis. In Section
\ref{sect:impulsive}, we analyse the impulsive phase of the flare; while in
Section \ref{sect:gradual}, we do the same for the gradual phase. Finally, 
we give our concluding remarks.

\section{Observations and principal characteristics of the event}
\label{sect:data}

\begin{figure}
\centerline{\resizebox{10cm}{!}{\includegraphics{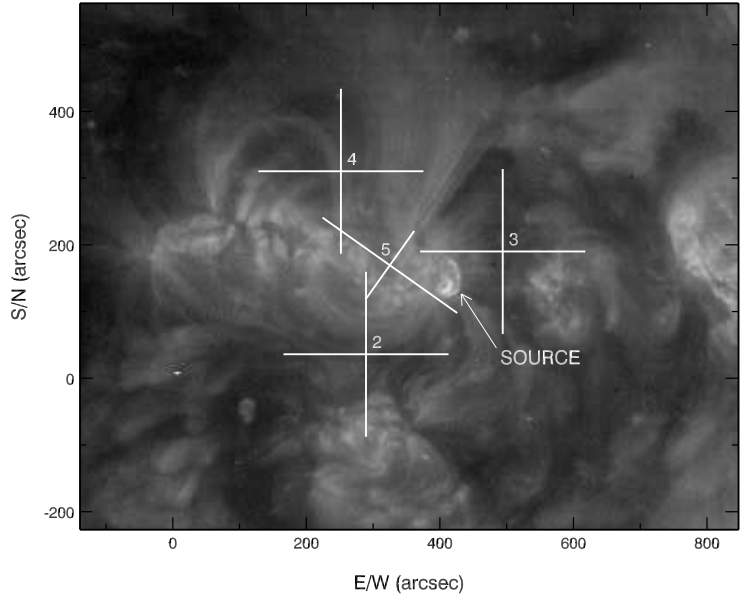}}}
\caption{A 195 \AA\ EIT image of the flaring area at 11:36:42 UT.  North 
is up and east to left. The labelled crosses represent the SST beams:
  2, 3 and 4 are 212 GHz receivers, while 5 is a 405 GHz receiver. Only
  beams 3 and 5 detected the burst. \label{fig:euv}}
\end{figure}

The SST \citep{Kaufmannetal:2001} is a radome enclosed 1.5 m single
dish antenna with room temperature receivers in a focal array: 4 at
212 GHz and 2 at 405 GHz, with 5 ms time resolution. On 30 October, 
2004, SST was tracking AR 10691 (N13 W18) with its beam 5. In
Figure \ref{fig:euv}, we sketch the position of SST beams with
crosses on a 195 \AA \ EUV image of EIT \citep{Delaboudiniereetal:1995}; 
the size of the crosses is equal to the HPBW.  While beams 2, 3 and 4 
(212 GHz) are round, beam 5 (405 GHz) has an elliptical shape, with 
minor and major axes in the directions shown in the figure.  Because of 
the high atmospheric attenuation only two beams observed this event, 
one at 212 GHz (beam 3) and one at 405 GHz (beam 5). Beams 1 (212 GHz)
and 6 (405 GHz) are not shown in the figure. Since only two beams 
observed the event, we cannot use the multibeam technique 
\citep{Costaetal:1995} to locate the centroid of emission.\\  

Assuming that the radio source is coincident with the EUV bright loop 
(see Fig.~\ref{fig:euv}), we correct the observed antenna temperature 
at 212 and 405 GHz for the misalignment. During calibration the antenna
temperatures are also corrected for atmospheric attenuation.  Finally,
the corrected antenna temperatures are converted to flux densities, $F$,
using the well known formula: $F = 2 k_b T_a / A_e$, where $k_b$ is
the Boltzmann constant, $T_a$ the corrected antenna temperature and
$A_e$ the effective area of the antenna. We estimate a 40\%
uncertainty for the whole process.\\

\begin{figure}
\centerline{\resizebox{10cm}{!}{\includegraphics{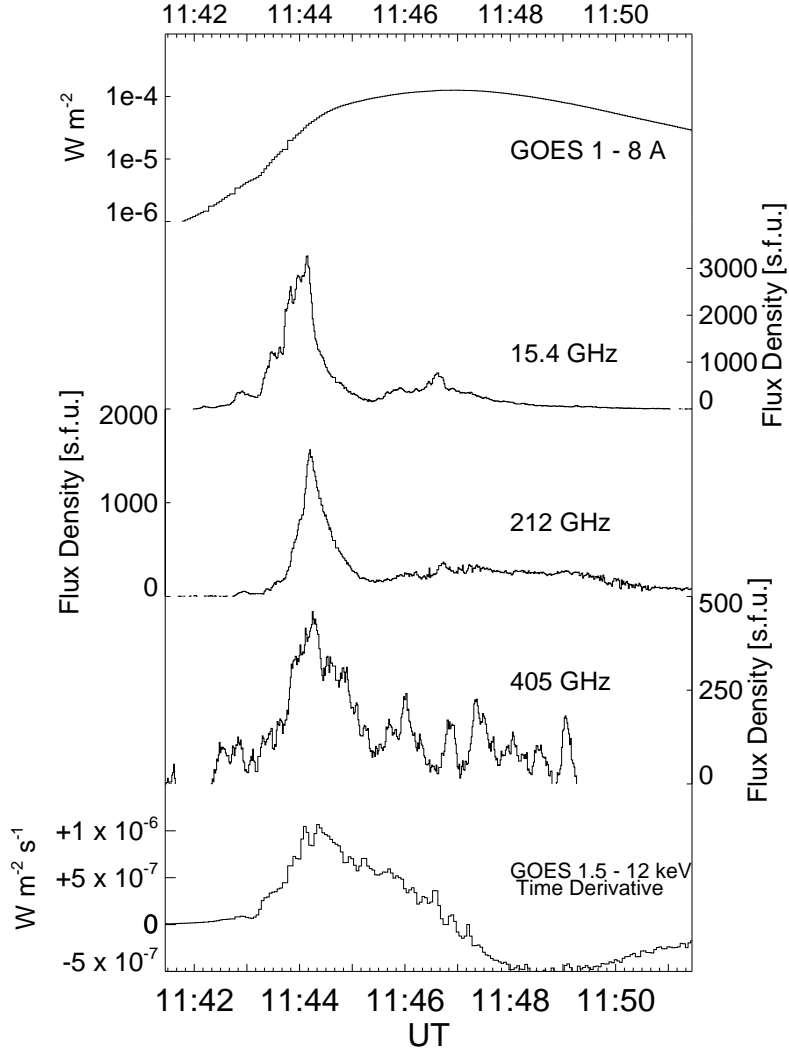}}}
\caption{Time profiles at selected wavelengths.  GOES data have 3 seconds
  time resolution, radio data are integrated to 1
  second. \label{fig:profiles}}
\end{figure}

Data at microwave frequencies with 1 second time resolution were
obtained by the Radio Solar Telescope Network 
\citep[RSTN,][]{Guidiceetal:1981}. Figure \ref{fig:profiles} shows the 
time profiles of the event at different wavelengths.  At 15.4 GHz, we see 
that the burst starts at 11:42:40 UT as a small enhancement that 
evolves into the impulsive phase with a maximum flux density around 
3300 s.f.u. at 11:44:09 UT. The increasing phase is characterized 
by short emissions that can be attributed to different particle 
injections. The decreasing phase is smooth, with no bursts, reaching 
the minimum value (close to the pre-flare level) at 11:45:25 UT. 
Afterward, a post burst increase is observed until approximately 
11:51 UT.  This gradual phase has also some bursting episodes 
characteristic of particle injections.  \\

Although the 212 GHz time profile is in general smoother than that 
at 15.4 GHz, most of the features observed at 15.4 GHz are also present, 
with smaller amplitude, at 212 GHz. The starting time of the burst at 
212 GHz coincides with that at 15.4 GHz (11:42:40 UT), but the peak time 
is delayed by 3 seconds (11:44:12 UT at 212 GHz). The gradual phase is also 
present with sudden changes, probably, of instrumental origin. The 405 GHz 
time profile during the rising phase is closer to the 15.4 GHz emission 
than to the 212 GHz, is in general delayed in 6 seconds. The decaying phase 
is more extended than in the other frequencies and shows the presence of 
longer features, some of these features could have an atmospheric origin, 
and do not allow us to clearly recognize the gradual phase. 

Unfortunately, no HXR data are 
available for this event.  Instead, we have used the time derivative 
of  GOES 1.5 -- 12 keV soft X-ray (SXR) channel with 3 seconds temporal 
resolution to compare with the radio flux density.  The sensitivity 
of GOES is better than $5 \ 10^{-8}$ W m$^{-2}$, taking this value as the
flux uncertainty, the flux time derivative has 
a mean uncertainty better than $\sqrt{2} \ 5 \ 10^{-8}/3$ W m$^{-2}$ s
$^{-1}\sim 3 \ 10^{-8}$ W m$^{-2}$ s$^{-1}$. The time derivative of GOES 
channel 3 (bottom curve in Fig.~\ref{fig:profiles}) mimics very well 
the radio time profiles during the rising phase and is coincident in time, 
but has two clear maxima, the first peaking in coincidence with the 15.4 
GHz emission. We will show later that the second GOES time derivative peak, 
still during the rising phase, is also present at the other frequencies, 
but during their decaying phase. Finally, the decaying phase in this curve 
lasts longer than the others and has some bursting episodes.\\

\begin{figure}
\centerline{\resizebox{9cm}{!}{\includegraphics{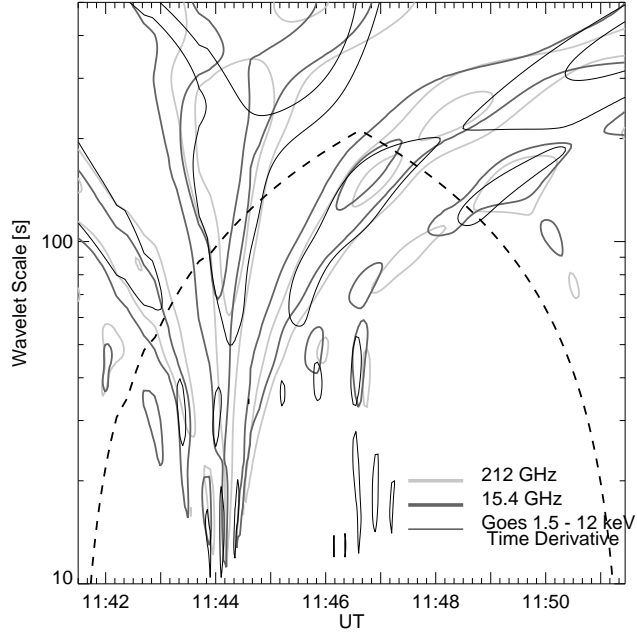}}}
\caption{ Wavelet transform of the flux density at 15.4 and 212 GHz and
GOES time derivative represented in a 2D graph. Contour levels are: $10^{-7}, \
10^{-6}$ W m$^{-2}$ s$^{-1}$ (GOES time derivative); 75, 750 s.f.u. (212
GHz) and 185, 1850 s.f.u. (15.4 GHz).\label{fig:wavelet}}
\end{figure}

We use the wavelet transform \citep[see e.g.][]{Daubechies:1992} to compare
the time profiles. Wavelets are very useful to discriminate the different
time scales in a time series.  We have integrated our data to the lowest
time resolution, namely 3 seconds and applied the wavelet transform
implemented by \citet{TorrenceCompo:1998} based on a Morlet mother wavelet
with wavenumber 6.   In Figure \ref{fig:wavelet} each contour
represents a structure resolved by the transform with a characteristic
scale in the ordinates occurring at the time in the abscissas.  We note
that for each frequency the minimum contour level plotted is much greater
than the signal uncertainty.  We have overplotted the density flux at 15.4
GHz (dark thick curve), 212 GHz (light thick curve) and the time derivative
of the GOES flux (black thin curve).  The dashed curve represents the {\em
cone-of-influence} of the transformation, i.e., where the boundaries do not
affect the result.  We observe that near peak time (11:44 UT) there are
three groups of structures with scales around 10--20 seconds.  The first
one, a little before 11:44 UT, is observed at the three frequencies with no
appreciable delay between them. The second one, just after 11:44 UT is
present also at the three frequencies. The last one ($\sim$ 11:44:30 UT) is
observed at 15.4 GHz and GOES time derivative.  At 212 GHz, this structure
appears only after 11:44:40 UT.  Other structures with longer duration are
also observed for the 40 seconds scale, around 11:45:50 UT and around
11:46:30 UT. In general, the wavelet transforms are similar for the
different frequencies, including the coincidence in time of the
characteristic features at the different scales.  Time profiles do not
reflect this in a more evident way because the relative importance of the
pulses are different.  \\

\section{The impulsive phase}
\label{sect:impulsive}
\begin{figure}
\center
  \hbox{
      \includegraphics[width=.5\textwidth]{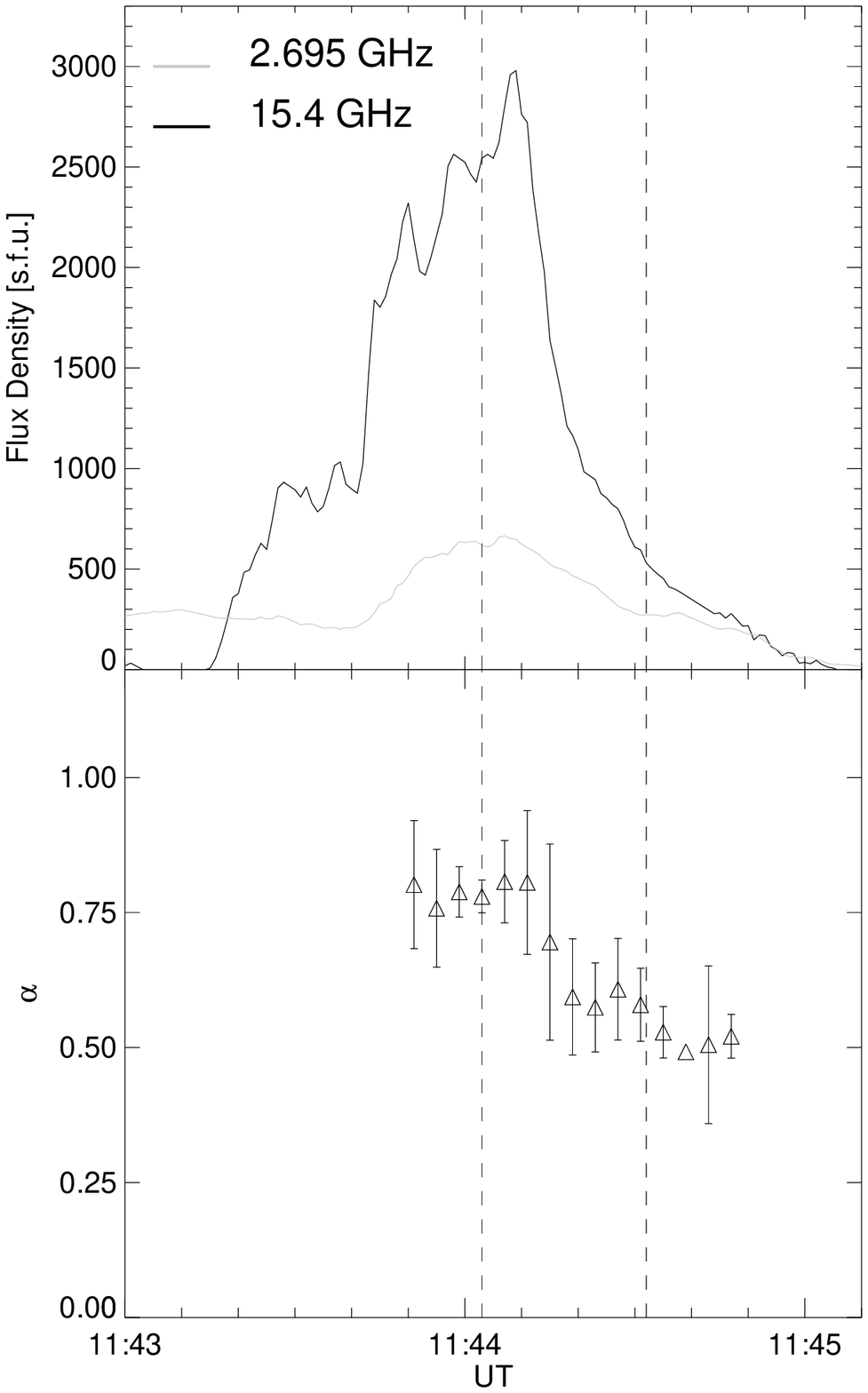}
  \hfill
    \includegraphics[width=.5\textwidth]{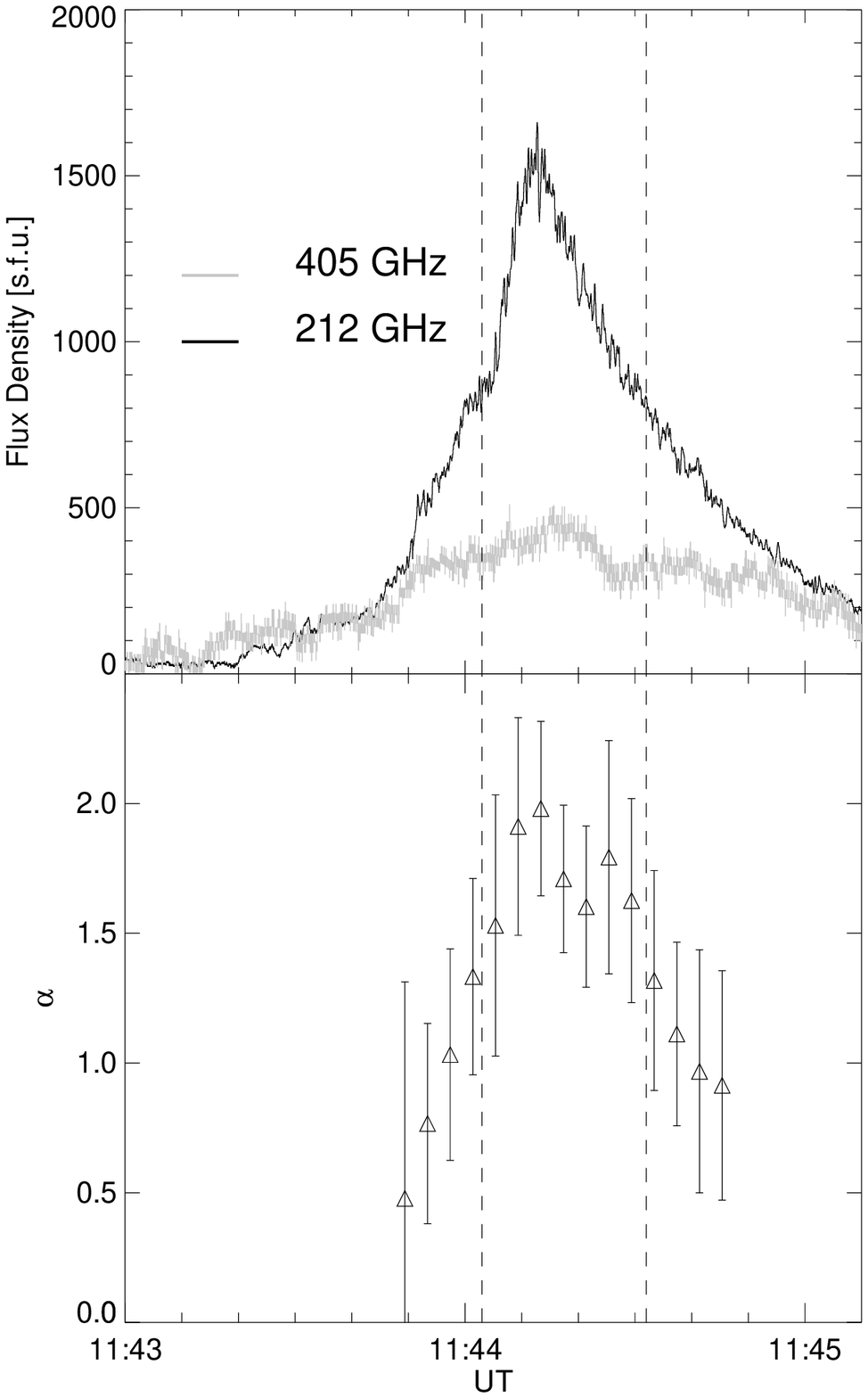}
    }
  \caption{Left panel: 1 s time profiles of 2.695 and 15.4 GHz flux densities
  from RSTN observations (above) and temporal evolution of the optically
  thick spectral index $\alpha$ obtained from the 2.695, 4.995, 8.8 and 
  15.4 GHz flux densities (below). Right panel: 40 ms time profiles of 405 
  and 212 GHz flux densities deduced from SST observations (above) and 
  temporal evolution of the optically thin spectral index $\alpha$ obtained 
  from the 405 and 212 GHz flux densities ratios (below).}
  \label{fig:alpha}
\end{figure}

 Figure~\ref{fig:alpha} (left panel) depicts the time evolution of the 
2.695 and 15.4 GHz emissions at the top, together with the temporal evolution 
of the optically thick spectral index at the bottom. This index is determined 
from the flux densities at 2.695, 4.995, 8.8 and 15.4 GHz. The index is 
lower than 1 all along the impulsive phase. The spectral index of the 
optically thick gyrosynchrotron spectrum originated in a homogeneous 
source is between 2.5 and 2.9 \citep{Dulk:1985}, independently of the 
electron index. Therefore, our result is an indication of the presence of 
an inhomogeneous source.\\

The 40 ms integrated time evolution of the flux density at 212 and 405
GHz during the impulsive phase is shown in the right panel of Figure
\ref{fig:alpha} (above).  The bottom of this panel corresponds to the temporal 
evolution of the  optically thin spectral index deduced from these 
flux densities, the only two frequencies that are optically thin in our 
data.  During the central part of the impulsive phase (between the dashed 
lines in the right panel of the Fig.~\ref{fig:alpha}), the index is close to 
constant, with gradual transitions at the beginning and end. Therefore, we can 
characterize the peak time with only one electron population. Using 
Dulk's formula \citep{Dulk:1985}, we can estimate the electron energy index as
$\delta\approx3$.

 The low spectral resolution of our radio data does not allow us 
to obtain a unique theoretical fit of the spectrum, i.e. different sets 
of parameters (magnetic field strength, viewing angle, source size, 
lower and higher cutoff energies, etc.) give similar results for the 
optically thin region; while, the optically thick region is never well 
fitted, as expected when an inhomogeneous source is present.

\section{The 212 GHz gradual phase}
\label{sect:gradual}

\begin{figure}
\centerline{\resizebox{9cm}{!}{\includegraphics{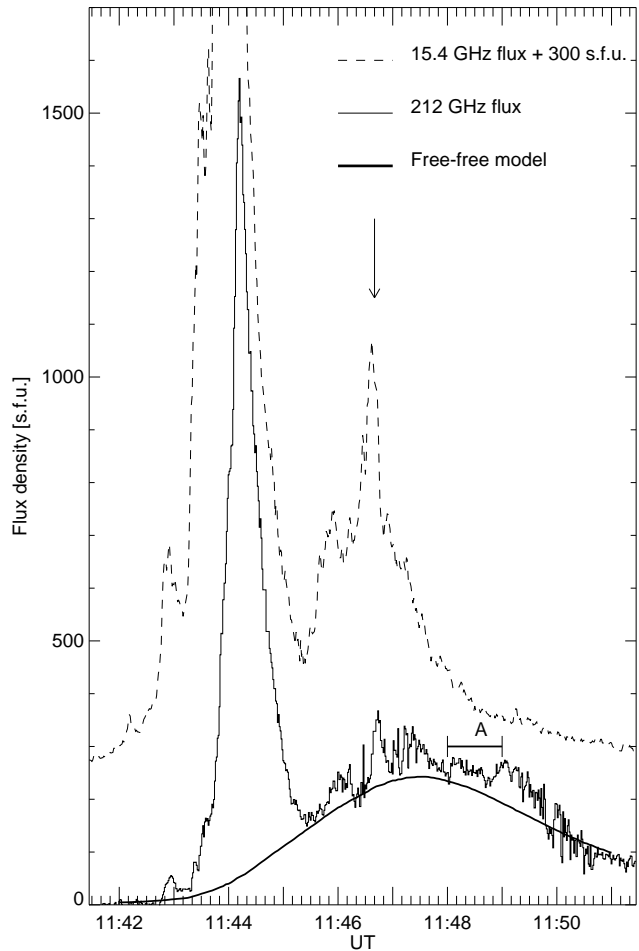}}}
\caption{The 212 GHz gradual phase.  After the burst ($\sim$ 11:45:30
UT), a gradual increase of the flux is observed at 212 GHz lasting
until 11:51:30 UT. During the gradual phase, some bursting episodes
are seen. In contrast, after 11:45:30 UT, the microwave flux, here
represented by the shifted 15.4 GHz (dashed line), shows a second
burst composed of many pulses.  The most intense one is also observed
at 212 GHz. Along interval {\bf A} we integrate the flux density at
the different frequencies to obtain one spectrum
(Fig. \ref{fig:ff-spec}).  The continuous thick line represents
the expected emission of a thermal bremsstrahlung. \label{fig:gradual}}
\end{figure}

The gradual phase is observed very clearly at 212 GHz
(Fig. \ref{fig:gradual}).  At lower frequencies, after the main burst,
new bursting episodes of gyroemission origin are seen.  
Figure~\ref{fig:gradual} shows the 15.4 GHz flux density, arbitrarily
shifted by 300 s.f.u., to clarify the plot; the late bursting episodes 
are evident in this figure. The strongest 15.4 GHz feature is also
observed at 212 GHz (arrow in Fig.~\ref{fig:gradual}). Variations in 
the atmospheric attenuation and/or emission are relatively more intense 
at 405 GHz than at 212 GHz, and thus they modulate the
background making the free-free emission difficult to observe. \\

\begin{figure}
\centerline{\resizebox{9cm}{!}{\includegraphics{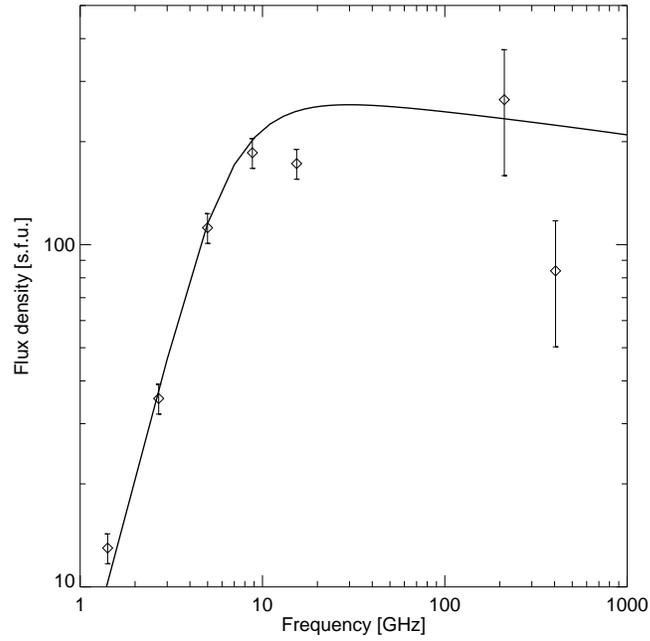}}}
\caption{Spectrum of the flux density during the interval 11:48 --
11:49 UT.  The solid curve represents the expected emission of an
isothermal bremsstrahlung source. \label{fig:ff-spec}}
\end{figure}

\begin{figure}
\centerline{\resizebox{9cm}{!}{\includegraphics{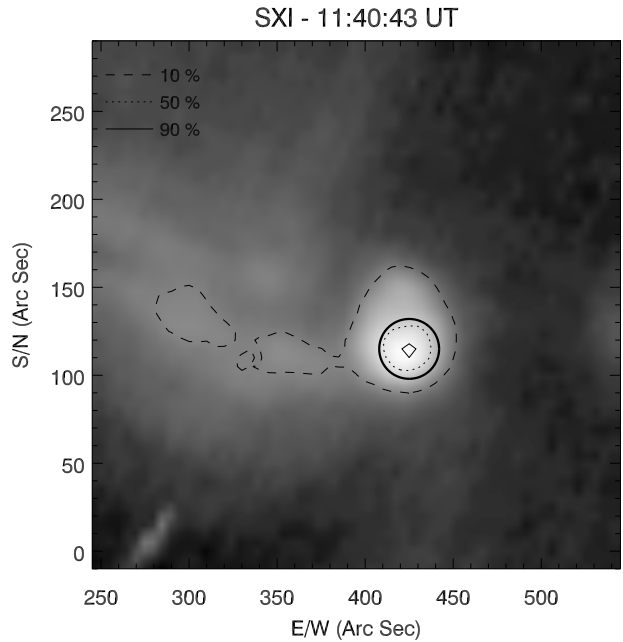}}}
\caption{A soft X-ray image taken by SXI of the flaring area just
before the maximum of the flare.  The contours represent the emission
levels of the thermal source.  The circle corresponds to the source
of our isothermal bremsstrahlung model. \label{fig:sxi}}
\end{figure}

To test the thermal bremsstrahlung hypothesis we have modelled the
observed emission as an isothermal homogeneous bremsstrahlung source.
We integrate the flux densities at different frequencies in the 1
minute interval shown in Figure \ref{fig:gradual} with a label 'A' to
obtain the gradual spectrum.  We have not tried to subtract the contribution
from other emission mechanisms, like gyrosynchrotron; therefore, we choose 
an interval that does not show strong gyroemission. To determine the size
of the source we have used an image from the Solar X-ray Imager 
\citep[GOES/SXI,][]{Hilletal:2005,Pizzoetal:2005}  
(Fig.~\ref{fig:sxi}), considering the emitting area above 30\% of the
maximum, which yields a source size of the order of 40''.  The emission
measure and temperature of the source are determined using the 
Chianti-based model with coronal abundances and the response of GOES 
detectors.  We fit the expected emission to the observed data, having 
as free parameters the source size and the source depth and we include 
a multiplicative factor $\zeta$. The time profile of the gradual 
emission at 212 GHz must be fitted simultaneously. We have obtained the 
best fit with a source size of 28'', $8.4 \ 10^8$ cm depth and  
$\zeta=7.5$. Figures \ref{fig:gradual} and \ref{fig:ff-spec} show the 
results.  The temporal evolution of the model is remarkably similar to 
the gradual phase of the 212 GHz emission. Moreover, the modelled 
spectrum follows the observed one. \\

The homogeneous and isothermal thermal bremsstrahlung flux density, 
$F(\nu)$, is:
\begin{equation}
F(\nu) = \zeta \ \frac{2 \ k_B \ T \ \nu^2}{c^2} \left ( 1 - 
         \exp{(-\kappa_\nu H)} \right ) \Omega \ , 
\label{eq:bremss}
\end{equation}
where $\kappa_\nu$ is the absorption coefficient, $H$ is the depth of
the source and $\Omega$ the source solid angle, while $T$ represents
the source temperature and $\zeta$ is our multiplicative factor. The
absorption coefficient $\kappa_\nu$ can be approximated by
\citep{Dulk:1985}:
\begin{equation}
\kappa_\nu \simeq \frac{0.2 \ N_{med}^2}{T^{3/2} \nu^{2}} \ 
      \mathrm{cm}^{-1} \ ,
\end{equation}
with $N_{med}$ the thermal plasma density. In the optically thick
regime, where $\tau_\nu = \kappa_\nu \ H >> 1$, Eq. \ref{eq:bremss}
tends to:
\begin{equation}
F(\nu) \simeq \zeta \frac{2 \ k_B \ T \ \nu^2}{c^2} \Omega \ .
\label{eq:ap-bremss}
\end{equation}
Therefore, if we want to explain the observed spectrum, the product $T
\Omega$ of the microwave source must be a factor 7.5 larger. In the
optically thin regime, $\tau_\nu << 1$, Eq. \ref{eq:bremss} turns out to be:
\begin{eqnarray}
F(\nu)&\simeq& \zeta \frac{2 \ k_B \ T \ \nu^2}{c^2}  \ \ 
       \frac{0.2 \ N_{med}^2 \ H}{\nu^2 \ T^{3/2}} \Omega \nonumber \\
      &\simeq& \zeta \frac{0.4 \ k_B}{c^2 \ R^2} \left (\frac{EM}{\sqrt{T}} 
\right ) \ ,
               \label{eq:of-bremss}
\end{eqnarray}
where $EM = N_{med}^2 V$ is the emission measure, $V$ the volume of
the source and $R$ is the Sun - Earth distance. From
Eq.~\ref{eq:of-bremss} we can infer that the product $EM/\sqrt{T}$
must be a factor 7.5 larger; i.e. either the EM is larger or
the temperature is lower, or both things are true.  Therefore, considering the 
optically thick and optically thin regimes, we conclude that the source is 
neither homogeneous nor isothermal.  Similar results for the gradual phase 
of other events were obtained by \citet{Luthietal:2004b} at submillimeter 
wavelengths, and by \citet{Chertoketal:1995} and \citet{Suietal:2005} in 
microwaves.

\section{Conclusions}
\label{sect:ending}

In this work we have presented a multiwavelength analysis of a GOES X1
class flare, including radio data, SXR and EUV images. Radio observations 
cover the range between 1.4 and 405 GHz, but we have no observations
between 15.4 and 212 GHz. At 212 GHz, the event has two
distinct phases: impulsive and gradual. At microwave frequencies,
there is a second impulsive phase whose emission at submillimeter
frequencies is barely observed.\\

  We find that the  optically thin spectral index $\alpha$ is almost
constant during the central part of the impulsive phase, afterwards there
is a hardening.  A transition to free-free emission, that in the optically
thin regime has an almost null index, can explain the change of $\alpha$.\\

The gradual phase observed at 212 GHz can be explained by brems\-strahl\-ung
from a thermal source. The similarity of the temporal evolution of the 
emission coming from an isothermal brems\-strah\-lung source (with $EM$ and 
$T$ deduced from SXR data) and the flux density at 212 GHz, sustains this
hypothesis.  Nonetheless, either the $EM$ must be higher or $T$ must
be smaller to fit the observations.  On the contrary, to
fit to the microwave observations, the product $T\Omega$ must be
larger.  Therefore, the thermal source cannot be homogeneous. \\

The use of submillimeter frequencies in the analysis of burst
phenomena opens new windows to understand the earlier stages of a
flare at high energies, which are hard to observe with HXR detectors,
and the late stages of the flare, where the thermal bremsstrahlung is
normally optically thin and more intense than at microwave frequencies. \\

{\em Acknowledgements}.This research was partially supported by the
Brazilian agency FAPESP (contract 99/06126-7) and by the Argentinean
grants: UBACyT X329 (UBA), PICT 12187 (ANPCyT), PIPs 6220 and 6266
(CONICET). G.D.C. is a fellow of ANPCyT. We acknowledge the GOES/SXI team 
for SXI data,  and the SoHO/EIT consortia for EUV data. SOHO is a 
project of international cooperation of ESA and NASA. The RSTN data originate 
at several United States Air Force monitoring stations.

\bibliography{cristiani2007a}
\bibliographystyle{elsart-harv}

\end{document}